\documentstyle[11pt]{article}
\begin{document}



\def\e{{\hat e}}
\def\ie{{$i.e.$}}
\def\ra{{\rightarrow}}
\def\a{{\alpha}}
\def\b{{\beta}}
\def\eps{{\epsilon}}
\def\n{{\eta}}
\def\g{\gamma}
\def\s{{\sigma}}
\def\r{{\rho}}
\def\z{{\zeta}}
\def\x{{\xi}}
\def\d{{\delta}}
\def\t{{\theta}}
\def\l{{\lambda}}
\def\ca{{\cal A}}
\def\cd{{\cal D}}
\def\ce{{\cal E}}
\def\cg{{\cal G}}
\def\co{{\cal O}}
\def\cn{{\cal N}}
\def\cs{{\cal S}}
\def\cz{{\cal Z}}
\def\cv{{\nu}}
\def\ck{{\cal K}}
\def\pr{{\partial}}
\def\prt{{\pr}_{\t}}
\def\tri{{\triangle}}
\def\na{{\nabla }}
\def\S{{\Sigma}}
\def\G{{\Gamma }}
\def\sp{\vspace{.1in}}
\def\hs{\hspace{.25in}}

\newcommand{\be}{\begin{equation}} \newcommand{\ee}{\end{equation}}
\newcommand{\bea}{\begin{eqnarray}}\newcommand{\eea}
{\end{eqnarray}}


\begin{titlepage}

\begin{flushright}
{{{\bf Nucl.Phys.B}(in press)}\\
{Goteborg-ITP-98-20}\\{hep-th/9812230}\\{December 24, 1998}}
\end{flushright}

\begin{center}
\baselineskip= 24 truept
\vspace{.5in}

{\Large \bf Path-Integral Formulation of Dirichlet String \\
in General Backgrounds}

\vspace{.4in}
{\large Supriya Kar\footnote{supriya@fy.chalmers.se}}

\end{center}
\begin{center}
\baselineskip= 18 truept

{\large \it Institute of Theoretical Physics \\
Goteborg University and Chalmers University of Technology \\
S-412 96 Goteborg, Sweden}

\end{center}

\vspace{.4in}

\baselineskip= 14 truept

\begin{center}
{\bf Abstract}
\end{center}
\vspace{.25in}

We investigate the dynamics of an arbitrary Dirichlet (D-) string in 
presence of general curved backgrounds following a path-integral formalism.
In particular, we consider the interaction of D-string with the massless
excitations of closed string in open bosonic string theory. The background
fields induce invariant curvatures on the D-string manifold and the 
extrinsic curvature can be seen to contain a divergence at the disk boundary.
The re-normalization of D-string coordinates, next to the leading order
in its derivative expansion, is performed to handle the divergence. Then we
obtain the generalized Dirac-Born-Infeld action representing the effective
dynamics of D-string in presence of the non-trivial backgrounds. On the
other hand, D-string acts as a source for the Ramond-Ramond two-form
which induces an additional (lower) form due to its coupling to the $U(1)$ 
gauge invariant fields on the D-string. These forms are reviewed in this
formalism for an arbitrary D-string and is encoded in the Wess-Zumino  
action. Quantization of the D-string collective coordinates, in 
the $U(1)$ gauge sector, is performed by taking into account the coupling 
to the lower form and the relevant features of D-string are analyzed in 
presence of the background fields.

\thispagestyle{empty}
\end{titlepage}

\baselineskip= 18 truept

\section{Introduction}

\hs
In the past few years, there have been remarkable progress in the
understanding of perturbatively different formulations of string theory
\cite{witten}. It is known that the Neveu-Schwarz (NS)-NS and Ramond-Ramond
(RR) background potentials have different nature in perturbation theory.
For instance, perturbative string states are charged with respect to the
NS-NS background field and are neutral under that of RR field. However
under string duality symmetry, NS-NS string states can be mapped to RR
charge states \cite{hulltown,witten,schwarz}. 
At the same time, there are several
other attempts to understand the non-perturbative insights in string and 
Yang-Mills theories. In most of these investigations, the D-branes 
\cite{polchinski, dailp} have played a significant role since it 
is a carrier of the elementary charges of the RR field.
As a consequence, a systematic treatment of 
non-perturbative effects is made possible by extending the world-sheet 
tool known for perturbative string states. In this context, some of the reviews 
\cite{pcj,bachas} can be found in the literature.

\sp
\hs
By now, it has been well established that D-branes 
are the RR sources required by the string duality \cite{polchinski}.
This in turn implies that the D-branes are inherent in type II superstring 
theories. Since type I string theory can be obtained by the orientifold
projection of the type IIB string theory, some of the D-branes also exist
in type I string theory. It is known that,
the D-brane tension arises from the disk
amplitude and is naturally associated with the inverse of closed string 
coupling which turns out to be that for branes with RR charges. There are 
massless open string states propagating on the D-brane and
they correspond to the collective coordinates for the transverse fluctuations
of the D-brane. The interactions of D-branes 
\cite{klebanov,cbachas,callank,garousi,hashimoto,hkazama,billo,kazama,
arfaei,karkazama} via
closed string exchange to describe their long range behavior have been
extensively studied. 
In this context, string tensions have been derived for strings carrying
both NS-NS and RR charges \cite{schwarz} and are interpreted as the bound
states \cite{witbound,miaoli,asen,myers} of D- and F-strings.
The supersymmetric extension of D-brane dynamics
in arbitrary backgrounds has also been investigated 
\cite{cederwall,cederwall2,schwarza,hatsuda,garousi2,kimura}.

\sp
\hs
In this paper, we extend the path-integral formalism, developed recently
\cite{karkazama} in collaboration with Y. Kazama, to incorporate the
general curved nature of backgrounds. In this formalism,
the D-string effective dynamics can be naturally taken into account, starting
from a D-string configuration in open 
string theory. By path-integrating out
the fundamental (F-) string modes one obtains the quantum dynamics of 
D-string in this formalism. In our earlier work \cite{karkazama}, we
have computed the renormalized vertex operator for a D-string in presence of
$U(1)$ gauge field of type I bosonic string theory. In addition, an attempt
was addressed to compute the quantum amplitude for the scattering between
elementary excitations of D- and F-strings. 

\sp
\hs
On the other hand, in the present work
we consider an arbitrary D-string in presence of 
massless closed string  
background fields, namely: metric, dilaton and the Kalb-Ramond (KR) 
antisymmetric
two-form potential, in open bosonic string theory. The $U(1)$ gauge field
on the D-string is a requirement for a consistent description of D-string in
presence of KR two-form. In fact, such a gauge field can also be seen as 
a part of the background $U(1)$ gauge field in open string theory. The
gauge field is attached to the ends of the open string and is 
responsible for the dynamics of D-string. In addition, D-string is
also charged due to the RR forms and these forms are completely independent
of the above backgrounds as mentioned in a previous paragraph. In an 
alternate description for the interaction of
D-string, the above background fields correspond to that of NS-NS sector in
type IIB string theory where D-string is a source of the appropriate RR forms.
These forms involve spin 
fields and are defined as the polarization in the 
corresponding vertex operator.
This complicates the non-linear sigma model coupling to the RR forms.
Thus a priori, it seems beyond our
scope to formulate the path-integral by taking into account the RR couplings.
Nevertheless, through an $SL(2,Z)$ string duality in the NS-NS sector of 
type IIB superstring theory, one can generate the required
differential form in the RR sector giving rise to the non-perturbative 
description to the D-string. Thus by considering the closed string background
fields of bosonic theory as that of the NS-NS sector {\footnote{In a different
context, we have used
a similar technique, in an earlier work \cite{kks}.}} 
of type IIB superstring,
one arrives at an alternative description of D-string in the bosonic
sector of type IIB superstring theory. Since in presence of the KR two-form
the D-string carries a lower-form of RR charge \cite{douglas}, a systematic
treatment of NS-NS and RR sectors of type IIB string theory is crucial to
discuss the quantum dynamics of D-string.

\sp
\hs
With the input, we compute the amplitude defining the effective action
for a D-string in presence of the closed string
background fields in open bosonic string theory. 
In this formalism,{\footnote{
Previously, this formalism has also been developed to study the scattering
of a D-particle with F-string states \cite{hkazama} and the
scattering among two D-particles \cite{kazama}.}} the
computation in the bulk is exact ( in $\a'$) even in presence of non-trivial
background fields and an approximation has been made only at the boundary.
Finally, we obtain the generalized Dirac-Born-Infeld (DBI) action \cite{leigh}
along with the Wess-Zumino (WZ) action describing the
interaction of D-string with the massless states, in the bosonic sector,
of type IIB string theory. Then we present the covariant quantization of
D-string collective coordinates in the $U(1)$ gauge sector to obtain the
quantum dynamics of D-string.
It is shown that the background fields can naturally lead to
various interesting aspects of D-string. To be specific, we analyze the
background fields to obtain the bound states of D- and F-string, the boost
and finally the D-instanton like configurations fixed with the fluctuations
of D-string. One of the strong motivation behind
the present work is to understand various alternative configurations of
D-string due to the presence of the general background fields in type IIB
superstring theory.

\sp
\hs
We plan to present the paper as follows. In section 2, we present a discussion
on the interactions of D-string with the massless excitations in the NS-NS and 
RR sectors of type IIB superstring in the path-integral formalism. In 
particular, we describe the boundary condition on the open string coordinates
in section 2.1, NS-NS and RR couplings to D-string in section 2.2, the
path-integral formulation for D-string dynamics in section 2.3 and in 2.4, we 
discuss the approximation used at the boundary leading to an orthogonal
moving frame. Section 3 deals with the computation of path-integrals involved
in the process. In section 3.1, we separate out the string zero modes, in 
section 3.2 
boundary conditions are discussed, then the path-integrals are evaluated
in the bulk and boundary respectively in sections 3.3 and 3.4. Section 4
essentially deals with some of the relevant configurations of D-string.
We present the quantization in section 4.1, bound state in 4.2, boost in
section 4.3 and fluctuations of D-string in 4.4. Finally we conclude with
discussions in section 5.

\section{Path-Integral Formalism}

\subsection{Curved closed string backgrounds}

Let us begin with the closed string massless background fields 
in open bosonic string theory. The arbitrary 
background fields are: the metric $G_{\mu\nu}$,
Kalb-Ramond (antisymmetric) two-form potential $B_{\mu\nu}$ and a dilaton 
$\Phi$ in space-time for ($\mu ,\nu = 0,1,\dots , 25$). 
These background fields can be
neatly coded in the NS-NS sector of a type IIB superstring which in its
bosonic sector also contains RR differential forms, namely :
an antisymmetric two-form $C_{\mu\nu}$, 
an axion or a zero-form $\chi$ and a four-form potential $C_{\mu\nu\r\s}$ 
with self dual field strength.
The theory is known to possess an exact
global $SL(2,Z)$ invariance under which the two-form potentials form a doublet
and the four-form is a singlet. The scalars can also be combined suitably and 
can be identified with the modular parameter of a torus. 

\sp
\hs
Now consider an arbitrary orientifold plane, with its space-time coordinates 
$f^{\mu}(t,\s )$ and the plane coordinates ($t,\s $), embedded in
the above non-trivial closed string background fields. By allowing 
open string fluctuations on the two dimensional orientifold plane one finds
dynamics on the plane and identifies it with an arbitrary D-string
world-sheet. Consider both ends of open strings to lie anywhere on the
orientifold plane to describe a D-string. Then the Lorentz
covariant condition at the boundary of the D-string
can be written as \cite{dailp}

\be
X^{\mu}({\t})\ = \ f^{\mu}\Big ( t({\t}) , {\s}({\t})\Big ) \ ,\label{dsdef}
\ee
where $X^{\mu}(z,\bar z)$ is the open string coordinates. We consider the 
topology of the
world-sheet of the open string to be an unit disk $\S$ and thus
parameterize its boundary $\pr\S$ with $\t$
for $(0\leq\t\leq2\pi)$. On the other hand, the ends of open string attached to
D-string is represented by the arbitrary functions
$t(\t )$ and $\s (\t )$.

\subsection{D-string charges}

With the above set up, let us analyze the dynamics in the bulk 
containing NS-NS and RR two-form potentials in more detail. Since the D-string
is in a general arbitrary backgrounds, it is useful to consider an uniform
description of the NS-NS and RR gauge symmetries.

\sp
\hs
First consider the KR two-form 
(NS-NS sector) in type IIB string theory. It couples in bulk to the
open string and the relevant part of the action can be given

\be
S(X,B) \ = - \ \ {i\over{4\pi\a'}} \int_{\S} d^2z 
\ \eps^{\bar a\bar b}B_{\mu\nu} \
\pr_{\bar a}X^{\mu}\pr_{\bar b}X^{\nu} \ , \label{antisymm}
\ee
where $\bar a, \bar b = 0, 1$ denote the world-sheet coordinates of F-string
and $\eps^{\bar a\bar b}$ is the antisymmetric tensor on the world-sheet.
Under a gauge transformation $\d B_{\mu\nu}  = \pr_{\mu}\Lambda_{\nu} -
\pr_{\nu}\Lambda_{\mu}$, where $\Lambda_{\mu}$ is an arbitrary one-form, the
action (\ref{antisymm}) is not invariant due to the presence of a disk boundary
$\pr\S$. In other words, the antisymmetric tensor induced on the D-string is
not invariant under the above gauge transformation where as the one, transverse
to the D-string remains invariant. This can be seen by writing the shift
introduced by the boundary term due to the gauge transformation $\d B$ as
\bea
\d S(X,B) &=& {i\over{4\pi\a'}} \int_{\pr\S} d\t \ \Lambda_{\mu} \pr_{\t}
X^{\mu} \ , \nonumber\\
&=&  {i\over{4\pi\a'}} \int_{\pr\S} d\t \ \Lambda_{a} \pr_{\t}
X^{a} \ , \label{shift}
\eea
where $\mu= (a,\a )$, $a =0,1$ 
denote the world-sheet coordinates ($t,\s$) for a D-string and $\a = 2,3,
\dots 25$ denote the transverse directions.
Then the change (\ref{shift}) is reduced to that on the D-string 
as the transverse part 
is taken care by the Dirichlet boundary conditions{\footnote{The
boundary conditions would be derived in a subsequent section 3.2.}}.
However, the gauge invariance can be restored by introducing an one-form
potential $A_{a}$
on the world-sheet of D-string, such that under 
the above gauge transformation of $B$,
$\d A_{a} = -\Lambda_{a}$. As a consequence, the gauge invariant  
field strength on the world-sheet of the D-string can
be written as $( B_{ab} + {\bar F}_{ab})$, 
where ${\bar F}_{ab}= 2\pi\a'\ F_{ab}$
in our notations. The gauge group
can be found to be a compact one, $i.e.\ U(1)$, 
rather than ${\bf R}$, which is in
fact can be seen to be a requirement for the consistency of type IIB string 
spectrum \cite{witbound}.
We shall see in the next section, this $U(1)$ gauge field $A_{\mu}$ is
essentially that of open string interacting at the disk
boundary{\footnote{Due to the Dirichlet boundary conditions, the
interaction of the gauge field can also be expressed in terms of a
space-time gauge field $A_{\mu}(X)$.}} . Since the
two-form potential $B_{\mu\nu}$ belongs to NS-NS sector of type IIB 
superstring, D-string automatically couples to NS-NS charge states.

\sp
\hs 
On the other hand, the RR two-form $C_{\mu\nu}$ also induces 
a corresponding two-form $C_{ab}$ on the world-sheet of a curved
D-string and can be
given by
\be
C_{ab} \ = C_{\mu\nu}(X)
\ \pr_{a}f^{\mu} \pr_{b} f^{\nu} \ .
\ee
Then, one may conclude that there are altogether two $U(1)$ 
gauge fields, one due to the KR two-form and the other one due
to the RR two-form 
respectively. This in turn leads to a $U(1)$ doublet under the 
$SL(2,Z)$. However a particular combination of two $U(1)$
gauge fields decouples leaving the other combination
of $U(1)$'s on the world-sheet of D-string \cite{townsend}. In other words,
apart form the gauge transformation of RR two-form{\footnote{In general, the
gauge transformation in the RR sector is different from that in the NS-NS 
sector.}}, it also undergoes a change
under the gauge transformation of KR two-form. Thus the complete gauge
transformation of the RR two-form requires a mixing of 
the RR two-form with the KR two-form potential in a gauge invariant way 
\cite{witbound,miaoli}. The invariant combination can be
taken into account to obtain the modified RR source term
on a three dimensional manifold describing the
WZ action whose boundary represents the world-sheet of an 
arbitrary D-string. It can be expressed as
\cite{douglas}

\be
\cs_{WZ}(f,C) \ = \ Q \ \int \ dt \ d\s \   
e^{(B \ + \ \bar F)}\ \wedge
C \ ,
\label{wz}
\ee 
where $Q$ denotes the charge density for a D-string. It can be expressed 
with respect to the dual of the field strength $H_3^{RR}\ ( =  d C_2)$ 
in RR sector:
$Q \ \equiv \ \int\ *H_3^{RR} $, where the integral is 
evaluated in the asymptotic region. While deriving the WZ action, the
boundary conditions describing the D-string have been taken into account.
The invariant combination in the exponential (\ref{wz}) is understood as 
the $(B+\bar F)_{ab} \wedge dy^{a} \wedge dy^{b}$. The appropriate RR forms are
schematically denoted as: $C$ on the world-sheet of D-string.
For instance, the induced two-form can be written as: $C_{ab}\
dy^{a} dy^{b}$. 
Thus the presence of KR two-form in the backgrounds induces
an $U(1)$ gauge field on the D-string world-sheet and can be seen to
carry a lower form of RR charge due its coupling (\ref{wz}).

\subsection{D-string in path-integral formalism}

\hs
Now we would like to consider the interaction of D-string with the massless
closed string background fields.
The perturbative excitations of the D-string are described by an
open string theory interacting with the closed strings in the bulk. 
In other words, we shall present a construction of the generalized 
DBI action for an arbitrary D-string in the path-integral formalism. 
The induced fields on the 
world-sheet of D-string give rise to the non-vanishing extrinsic curvatures
and thus responsible for the curved nature of D-string. 

\sp
\hs
We begin with arbitrary abelian gauge field $A_{\mu}(X)$ in space-time
coupled to the open bosonic string at the boundary. However, as mentioned
before in section 2.2, the boundary condition takes care of the transverse
components of the $U(1)$ gauge fields and one finds only the gauge field on the
world-sheet of D-string. Now,
in the open string language, we can encode the available informations for a
D-string configuration in the closed string backgrounds
by writing the Polyakov path-integral. The effective D-string dynamics can be
expressed in this formalism \cite{karkazama}:

\bea
{\cs }_{\rm eff}\Big ( f, A, B \Big )\  =
{1\over{g_s}} \int \cd X^{\mu}(z,\bar z )
\ \cd t({\t}) \ {\cd \s}({\t})&&\cdot {\d } \Big (\ X^{\mu}(\t) - f^{\mu} 
(t(\t) , {\s}(\t))\ \Big ) \nonumber\\
&&{}\qquad\qquad \cdot \exp \Big (-S[X,A,B ]\ \Big )\  , \label{path}
\eea
where $g_s= e^{\phi}$ is the closed string coupling constant. The  
$\d$-function in eq.(\ref{path}) ensures the Lorentz covariant condition and
can be seen to introduce Dirichlet boundary condition on some of 
the string modes.
$S[X,A,B]$ is the open string action coupled to the gauge field at the boundary
and interacting with the spacetime 
background fields: graviton, antisymmetric
tensor and the dilaton. In a conformal gauge, the world-sheet metric
of the F-string, $\g_{\bar a \bar b} = e^{2\b} 
\n_{\bar a \bar b}$. Then the open string action can be expressed \cite{callan}

\bea
S[X,A,B ]\ = \ {1\over{4\pi\a'}} && \int_{\S} d^2z \ \Big (\ G_{\mu\nu}(X)
{\pr_{\bar a}} X^{\mu}
{\pr^{\bar a}} X^{\nu} - \ i \ {\eps}^{\bar a\bar b} \ B_{\mu\nu}(X) 
{\pr_{\bar a}} X^{\mu} {\pr_{\bar b}} X^{\nu}  \nonumber \\
&& +\ \a' \ \na_{\mu}\Phi (X) \ 
\pr_{\bar a} X^{\mu} \ \pr^{\bar a}\b \Big ) \
+ i \ {\int}_{\pr {\S}} d\t \ G_{\mu\nu}(X) A^{\mu} (X) 
\ {\prt} X^{\nu} \ ,\label{action}
\eea
Note that the dilaton term appears as a
higher order in $\a'$ in comparison to the rest of the closed string background 
fields. The above action (\ref{action}) represents the coupling of the 
closed string fields to the open string in bulk and the
background gauge field at the boundary. In addition, there
is also an antisymmetric two-form potential in the RR sector. Following the
discussion in the previous section 2.1, one can consider the coupling of
the induced RR two-form potential on the world-sheet of D-string.
Then the dynamics of the D-string in type IIB theory can be described by the 
effective action

\be
S_{\rm D-string} \ = \ {\cs }_{\rm eff} (f,A, B) \ 
+ \ \cs_{WZ} (f,C)  \ .
\label{dstring} 
\ee
The second term in eq.(\ref{dstring}) corresponds to the WZ action (\ref{wz})
and accounts for the forms in the RR sector of type IIB string theory.
It is exact because of its non-perturbative nature. The first term in 
eq.(\ref{dstring}) describes the effective dynamics of D-string and is 
essentially associated with the closed string backgrounds. Now we shall
present the computation for the effective
dynamics ${\cs}_{\rm eff}$ of D-string.

\subsection{Approximation at the boundary}

\hs
In this section, we deal with the boundary interactions of D-string which is
essentially due to the fluctuations of the ends of the open string lying on
the world-sheet of D-string. These
boundary interactions can not be dealt exactly like that of bulk. Thus we
need to describe the boundary interaction with the help of an approximation
at the boundary. One such approximation preserving general coordinate
invariance is the geodesic normal coordinate expansion \cite{alvarez}. We
follow ref.\cite{karkazama} to re-write the expansion of 
$f^{\mu}(t(\t),\s(\t))$
order by order in the derivatives of the
D-string world-sheet coordinates around $f^{\mu}(t,\s )$. It can
be given

\bea
f^{\mu}\pmatrix {{t(\t),\s(\t) } \cr }&=& f^{\mu}\big ( {t,\s}\big ) \
+ \ {\pr }_a f^{\mu}\Big (t({\t }),\s ({\t })\Big ) {\z }^a ({\t }) 
\ + \ {1\over2} K^{\mu}_{ab}
{\z }^a({\t }) {\z }^b({\t }) \nonumber \\
&&{}\qquad\qquad\qquad\quad
+ \ {1\over{3!}} K^{\mu}_{abc} {\z }^a({\t }) 
{\z }^b(\t ) {\z }^c(\t ) \ + \ {\co}({\z }^4 ) \ , \label{geodesic}
\eea
where ${\z }^a({\t })$ are the geodesic normal coordinates for $a=(t,\s )$ 
and $K^{\mu}_{ab}(t,\s )$ and $K^{\mu}_{abc}(t, \s )$ are the extrinsic
curvatures of D-string manifold. They can be expressed in terms of the
projected space $P^{\mu\nu}$ normal to the world-sheet

\bea
K^{\mu}_{ab}  &=&  P^{\mu\nu}\ \pr_a\pr_b f_{\nu}\ , \nonumber \\
K^{\mu}_{abc}  &= & \pr_a\pr_b\pr_c f^{\mu} - \pr_a {\G^{\mu}}_{bc}
+2 {\G^d}_{ab} {\G^{\mu}}_{dc} \ ,\nonumber 
\eea
where $\G$ is the Christoffel connection. In fact, the normal projection
operator  $P^{\mu\nu}$ can be expressed in terms of a tangential 
projector $h^{\mu\nu}$:
\be
P^{\mu\nu} \ = \ G^{\mu\nu} \ + {B}^{\mu\nu} \ - \ h^{\mu\nu} \ .
\label{projection} \ee
Thus the closed string background fields can alternatively be described by the 
normal and tangential projected spaces.{\footnote{Similar treatment
in presence of background fields can be found in ref.\cite{dailp} in 
a different approach. There,
antisymmetric two-form has been treated as a vertex along with the gauge field
strength unlike the present case.}
The tangential space induces 
background fields in the D-string manifold and can be expressed 
\bea
h^{\mu\nu} \ &=& \ {\pmatrix { h^{ab} + B^{ab} }}
{\pr }_a f^{\mu}{\pr }_b f^{\nu} \ ,\\
h_{ab} \ &=& \ G_{\mu\nu}(X) \ {\pr }_a f^{\mu}{\pr }_b f^{\nu} \nonumber \\
{\rm and} \qquad  \ \ B_{ab} \ &=& 
\ B_{\mu\nu}(X)\ {\pr }_a f^{\mu}{\pr }_b f^{\nu} \ ,
\eea
where $h_{ab}$ and $B_{ab}$ are respectively the induced background fields:
metric and KR two-form on the D-string manifold. The expression  
(\ref{projection})
treats the background field $B_{\mu\nu}$
in the same footing as $G_{\mu\nu}$. 

\sp
\hs
Next, it can be seen that, to the leading order
in the expansion (\ref{geodesic}),  
the boundary of the disk is mapped on to a
point $(t,\s )$ on the D-string world-sheet and the sub-leading terms
denote the corrections around that point. Thus, we set up an orthonormal
moving frame{\footnote{Such a frame has
been discussed previously in refs.\cite{hkazama,kazama,karkazama} for 
a flat background metric.} 
at that point on the D-string in presence of the general backgrounds.
As we proceed, it can be seen that such an orthonormal frame simplifies our 
computation, by introducing a local inertial frame at the point $(t,\s )$
on the world-sheet of D-string. Then two of the basis
vectors, $\e^{\mu}_a, \, a=0,1 \ \equiv (t, \tilde\s )$, 
lie on the tangential plane with Neumann
boundary conditions and the rest, $\e^{\mu}_\alpha, \, \alpha = 2,3,\ldots, 25$
lie on a normal plane with Dirichlet boundary conditions. In general, these
orthonormal vectors ${\e^{\mu}_A}$,  
for $A = (a, \a )$ can be expressed
\bea
\e^{\mu}_A &=& \cn_{(A)} \ e^{\mu}_A \ , \nonumber \\
{\e^{\mu}}_0 &=& {{{\dot f}^{\mu}}\over{\sqrt{-h_{00}}}} \ ,\nonumber \\
{\e^{\mu}}_1 &=& {\sqrt{{h_{00}}\over{h}}} \ 
\Big ({f'}^{\mu} \ + \ \e^{\mu}_0
\e^{\nu}_0 \ {f'}_{\nu} \Big ) \ \equiv \ \cn_{(1)} \ 
\pr_{\tilde\s} f^{\mu} \ , 
\nonumber \\
G_{\mu\nu}\ \e^{\mu}_A \e^{\nu}_B &=& \n_{AB} \ , \nonumber\\
B_{\mu\nu} \ \e^{\mu}_A \e^{\nu}_B &=& \ \ce_{AB} \ , \nonumber \\
G^{\mu\nu} &=& {\e^{\mu}}_A {\e^{\nu}}_B \ \n^{AB} \ , \nonumber\\ 
B^{\mu\nu} &=& {\e^{\mu}}_A {\e^{\nu}}_B \ce^{AB} \ , \nonumber \\
h^{\mu\nu}  &=&  {\e^{\mu}}_a {\e^{\nu a}} 
\ + \ B^{ab} \ \e^{\mu}_a \e^{\nu}_b \ 
\nonumber\\
P^{\mu\nu} &=& {\e^{\mu}}_{\a} {\e^{\nu\a}} \ + \ B^{\a\b} \ \e^{\mu}_{\a}
\e^{\nu}_{\b} \ ,
\eea
where $\cn_{(0)} = {1\over{\sqrt{-h_{00}}}} $,
$\cn_{(1)} = {\sqrt{{h_{00}}\over{h}}}$,
$\cn_{(\a )} = ( 1,1,\dots  1 )$,
$h$ denotes $\det h_{ab}$, 
a dot and a prime on $f^{\mu}(t,\s )$ 
correspond to time $(t)$ and spatial $(\s )$
derivatives respectively. Here, 
$\n_{AB}$ and $\ce_{AB}$ are the flat
backgrounds representing the Minkowskian metric and the KR two-form 
respectively.
\section{Disk Amplitude}

\subsection{Non-zero modes of F-string}

Now for simplicity, the non-zero modes (say) $\x^{\mu}$ can be 
separated out from the 
string coordinates, $X^{\mu}(z, \bar z ) \ = \ x^{\mu} \ 
+ \ {\x}^{\mu} (z, \bar z )$, and in the orthonormal frame can be given by

\be
\x^{\mu}(z, \bar z )\ = \ \sum_A \e^{\mu}_A \ \r^A (z, \bar z ) \ ,
\ee
where $\r_a(z,\bar z ) $ and $\r_{\a}(z, \bar z ) $ correspond to the 
non-zero string modes in tangential
and transverse directions respectively.
Now it is straightforward to see that the non-trivial contribution
from the Lorentz covariant condition (\ref{dsdef}), 
expressed by the $\d$-function in eq.(\ref{path}), reduces to

$$\d \Big ( {\x }^{\mu} (\t ) - \pr_a f^{\mu}( t, \s )
{\z }^a (\t ) - \dots \ \Big ) \  .$$

\noindent Thus the non-zero modes of F-string are constrained by the collective
coordinates of D-string. On the other hand, the zero mode ($x^{\mu}$) 
is replaced by its
counterpart $f^{\mu}(t,\s)$ and does not contribute substantially apart
from a D-string volume factor in the path-integral formalism (\ref{path}).

\sp
\hs
The background fields can be
expanded around the zero-mode $x^{\mu}$ of the string $X^{\mu}$. 
Since WZ action is exact,
the relevant expansion of the background metric: 
$G_{\mu\nu}(x+\x )$, KR two-form: $B_{\mu\nu}(x+\x )$ and the dilaton
$\Phi(x+\x )$ 
around the zero mode $x^{\mu}$ are

\bea
G_{\mu\nu}(X) &=& \n_{\mu\nu} \ + \pr_{\l} G_{\mu\nu }\ \x^{\l}
\ + \ \dots 
\ , \nonumber\\
B_{\mu\nu}(X) &=& \ce_{\mu\nu} \ + \pr_{\l } B_{\mu\nu}\ \x^{\l} \ + \
\dots  \ ,
\nonumber\\
\Phi(X) &=& \phi \ + \ \pr_{\l}\Phi \ 
\x^{\l} \  + \ \dots  \ \ .
\eea
Now the $U(1)$ gauge field $A_{\mu}(X)$ can also be expanded around the
zero-mode $x^\mu$. Then the boundary interaction of the gauge field
becomes
\bea
i \ \int_{\pr\S} d\t \ A_{\mu} \Big ( x + \x \Big ) \ \pr_{\t} X^{\mu}
\ = &{{i}\over{2}}& F_{\mu\nu} \ 
{\int}_{\pr\S} d\t \ \x^{\mu} \pr_{\t} \x^{\nu} \nonumber \\
&& + \ {i\over3} \pr_{\l}
\ F_{\mu\nu} \int d\t \ \x^{\l} \x^{\mu} \ \pr_{\t}\x^{\nu} \ + \dots .
\eea
Now the action (\ref{action}) for the non-zero modes can be re-written 
\bea
S[\x ,B,A]\ = \ {1\over{4\pi\a'}} \int_{\S} && d^2z \Big (\ 
\Big [\ 
{\pr_{\bar a}} \x^{\mu}
{\pr_{\bar a}} \x_{\mu} - i \ {\eps}^{\bar a\bar b}\ {\ce}_{\mu\nu}\
{\pr_{\bar a}} \ \x^{\mu} {\pr_{\bar b}} \x^{\nu} \nonumber\\
&& + \ \pr_{\l} G_{\mu\nu}\ \x^{\l} \ {\pr_{\bar a}}\x^{\mu} 
\pr_{\bar a} \x^{\nu } \ - \ i\ \eps^{\bar a\bar b}
\ \pr_{\l }B_{\mu\nu }\ \x^{\l }\ {\pr_{\bar a}}\x^{\mu} 
\pr_{\bar b}\x^{\nu }\ \Big ] 
\nonumber\\
&& +\ \a' \ \pr_{\bar a}\b \ \Big [ \ 
\nabla_{\mu }\nabla_{\nu }\Phi \  
\x^{\nu } \pr_{\bar a} \x^{\mu} \ + \ \nabla_{\mu}\nabla_{\nu}\nabla_{\l }
\Phi \ \x^{\l } \x^{\nu}\pr_{\bar a}\x^{\mu }\  \Big ] 
\ \Big ) \nonumber \\
&+& {i\over2} F_{\mu\nu} {\int}_{\pr {\S}} d\t \ \x^{\mu}
\ {\prt} \x^{\nu} \ + \ {i\over3} \ \pr_{\l} F_{\mu\nu} \int d\t \ \x^{\l}
\x^{\mu} \prt \x^{\nu} + \dots \ ,\label{nonzero}
\eea
where dots denote the higher order terms in $\x$ containing the 
higher derivatives of metric $G_{\mu\nu}$, two-form $B_{\mu\nu}$ and the
dilaton in the bulk
and that of the $U(1)$ field strength $F_{\mu\nu}$ at the boundary.

\subsection{Boundary conditions}

\hs
In order to discuss the boundary conditions, we re-write the action 
(\ref{nonzero}) in terms of orthonormal coordinates and after
simplification, one obtains

\bea
S[\r,B,A] \ = \ {1\over{4\pi\a'}} \Big [ &-& \int_{\S} d^2z \ \Big (\ 
\r_A \pr^2 \r^A
\ - \ \a'\ \pr_{\bar a}\b \
\Big [\ 
\pr_A\pr_B\Phi \ \r^A \pr_{\bar a}\r^B\ + \ \dots \ \Big ] \ \Big ) \nonumber\\
&&{} \ +\ {\int}_{\pr\S} d\t \ \Big ( \ \r_A \pr_n \r^A 
\ + \ i\ \cn_{(A)} \cn_{(B)} 
\left (B_{AB} + {\bar F}_{AB} \ \right )\ \r^A
\pr_{\t} \r^B \ \Big ) \nonumber\\
&& {} \ \ \ \ + \ {{2i}\over3}\cn_{(B)}\cn_{(C)}
\ \pr_A {\bar F}_{BC} \int d\t \
\r^A \r^B \prt \r^C
\ + \ \dots  \ \Big ] \ ,\label{raction}
\eea
where $\pr_n$ denotes the exterior normal derivative, 
$B_{AB} \equiv B_{\mu\nu} e^{\mu}_A e^{\nu}_B, 
{\bar F}_{AB} \equiv 2\pi\a' \ F_{AB}$,
$\pr_A \equiv e^{\nu}_A \nabla_{\nu}$ and  
$A_B \equiv e^{\mu}_B A_{\mu}$. In the orthonormal frame, the derivative
corrections to the background 
metric and two-form vanishes and thus the treatment for these fields is
exact in bulk. Note that the KR antisymmetric potential has
appeared on the boundary (\ref{raction}) along with the gauge field interaction term, even though to start with, the interaction was expressed as an integral
in bulk. 

\sp
\hs
Since the dilaton term in eq.(\ref{raction}) is higher order in $\a'$ than
the rest of the background fields, the consistent set of boundary conditions
can be given by

\bea 
&&\pr_n \r_a (\t ) \ + \ \ {i\over{{\sqrt{- h}}}} \
\left ( B_{ab} + \bar F_{ab} \right ) \ \pr_{\t}\r^b(\t ) \ = \ 0
\nonumber \\
{\rm and }\qquad \qquad && \qquad\r_{\a}(\t ) \ = \ 0 \ , \label{bcond}
\eea 
where $(B_{ab}+ {\bar F}_{ab})$ is the invariant $U(1)$ field strength on the
world-sheet of D-string. The first expression in eq.(\ref{bcond}) is the
usual Neumann condition modified due to the presence of background fields . 
The second expression there
represents the Dirichlet conditions and defines the position of D-string in the transverse space.

\subsection{Path-integral in bulk}

\hs To evaluate the path-integral (\ref{path}) for the effective dynamics
of D-string, the action
(\ref{raction}) should be conformally invariant. Considering an off-shell
extension of the amplitude \cite{polyakov}, one can get rid of the dependence
on the F-string conformal factor $\b $.
In presence of gauge field, the
the suitable prescription becomes the gauge $\b = 1$ \cite{tseytlin}.

\sp
\hs
Now the effective action $\cs_{\rm eff}(f,A,B)$ in 
eq.(\ref{path}) needs to be evaluated
by path-integration in bulk as well as over the boundary fields. 
As a first step, we consider the path-integral in bulk along with
that of boundary over the non-zero string modes $\r_A(z,\bar z )$. 
The integration can be
performed uniformly in the bulk by re-writing the
boundary term for the non-zero modes appropriately in the bulk. Assuming an
adiabatic approximation for the background gauge field, the higher 
derivatives of the $U(1)$ gauge field strength ${\bar F}_{AB}$ can be ignored.
Then the path-integral becomes

\bea
I_{\r} \ = \ \int &&\cd\r_A \  \exp\Big ( -{1\over{4\pi\a'}}\int_{\S} d^2z
\ \Big [\ \pr_{\bar a}\r^A \pr^{\bar a}\r_A \
-\ \cn_{(A)}\cn_{(B)}\Big ( B_{AB}
+ {\bar F}_{AB} \Big ) \nonumber \\
&& \cdot \d (|z| - 1) \ \r^A\pr_z\r^B \Big ] \ \Big )
\ \cdot \exp \Big ( \ i \int_{\S} d^2z\ \d ( |z| - 1 )
\ \nu_a(z)\ \r^a(z) \ \Big ) \nonumber \\
&& \cdot \exp \Big ( -{1\over{3! \ \pi\a'}}
\cn_{(A)}\cn_{(B)}\ \pr_C {\bar F}_{AB} 
\int_{\S}\ d^2z \ 
\d\Big ( |z|-1\Big )\ \r^C(z) \ \r^A(z) \prt \r^B(z) \ \Big )\ ,\label{string}
\eea 
where $\d(|z| - 1)\ \nu_a(z)= \nu_a(\t )$ denote the auxiliary 
fields at the boundary.{\footnote{The Lagrange multiplier fields are due 
to the Lorentz covariant condition at the disk boundary.}} 

\sp
\hs
Being an orthonormal frame, the path-integral can be dealt separately
over the transverse and the longitudinal components. The path-integral
over the transverse components $\r_{\a}$ in eq.(\ref{string}) is a Gaussian
and becomes trivial. It does not contribute substantially to the 
partition function. On the other hand, the path-integral over the
longitudinal components $\r_a$ is non-trivial due to 
the presence of the $U(1)$ gauge invariant field strength and its derivatives.
However, it can be performed in presence of 
cubic interaction ( in $\r^a$)
as usual by introducing a source term. 
The Gaussian part of the integral is found to contain a non-trivial
Jacobian factor along with a path-integral expressed over the auxiliary fields
$\nu_a(\t)$. By using the Fourier mode expansion on a boundary circle
and performing the $\z-$function regularization \cite{tseytlin},
we obtain 

\be
{\rm Jacobian} \ = \ 
\left ( \ {{h \ +\ B \ + \ \bar F}\over{h}} \ \right )^{1\over2} \ ,
\ee
where $B + \bar F$ denotes the $\det (B + \bar F)_{ab}$ and 
thus $ h + B +\bar F$
is understood as $\det (h + B +\bar F )_{ab}$.

\sp

\noindent
Evaluating the cubic part along with the Gaussian part,
the result of path integration in bulk over the string coordinates 
$\r_A(z,z' )$ is given by

\bea
I_{\rm bulk} = {\sqrt {{h \ + \ B \ + \ {\bar F}}\over{h}}} &&
\cdot \exp \Big ( \ {{\a'}\over2} \int d\t \ d\t' \ \cv_a(\t ) \
G(\t , \t' ) \ \cv_b (\t' ) \ \n_{ab} \ \Big ) \nonumber\\
&& \cdot \exp \Big (\ {{-1}\over{3!\ \pi\a'}} 
\cn_{(a)}\cn_{(b)} \ \pr_a {\bar F}_{ab}\ 
\cdot I_3^b \ [ \ G(\t,\t)\ ] \ \Big ) \ , \label{bulk}
\eea
where $G(\t,\t')$ 
can be seen to be the Neumann propagator $G_{ab}(z,z')$ on the boundary of an
unit disk and the boundary integral due to the cubic interaction

\be 
I_3^b \ [\ G(\t,\t)\ ] \
= \ \int \ d\t \ G(\t, \t)\ \ \prt\r^b(\t ) \ .
\label{cubic} 
\ee
From eq.(\ref{bulk}) it is clear that, the interaction is that of boundary.
Thus we need to deal
with $G(\t,\t')$ instead of the propagator in the bulk. 
On the disk boundary $\pr\S$, the propagator satisfies
(\ref{bcond})

\be
\pr_n G(\t,\t' ) \ + \ {i\over{\sqrt{-h}}} \ 
\left (\ B_{ab} \ + \ {\bar F}_{ab}
\ \right ) \prt G(\t,\t') \ = \ 0 \ .
\ee
The Neumann propagator 
{\footnote{The propagator contains
background fields induced on the D-string, which is similar to the treatment
in ref.\cite{callan}, in the context of open string theory.}}
$G(\t,\t' )$ diverges as $\t' \ \ra \ \t$. One
can regularize the propagator \cite{tseytlin} by introducing a cut off 
parameter $\eps$ and obtains

\be
G(\t ,\t') \ = \ - 2\ h \left [h - \left (B + {\bar F}\right )\right ]^{-1} 
{\sum_{n=1}^{\infty}} {{e^{-\eps n}}\over{n}} \cos\ n (\t - \t') \ .
\ee
For a constant U(1) gauge invariant field strength 
$(B_{ab}+{\bar F}_{ab})$, the Neumann propagator is exact and the path-integral
becomes quadratic in string coordinates.
In the limit $\t \ \ra \ \t'$, the boundary propagator $G(\t,\t)$ 
becomes

\be
G_{aa}(\t,\t) \ = \ 2 \ \d_{aa}\  h \ 
\left [ h - \left ( B + \bar F \right )\right ]^{-1} \ln \eps \ .
\ee
Thus the cubic interaction term $I_3^b \ [G(\t,\t)\ ]$ in eq.(\ref{cubic})
contains a divergent term $G(\t,\t)$ and is independent of the
fluctuations (constant). However, an off-shell extension 
\cite{nappi} adds up to the non-constant part of the
gauge field strength ${\bar F}_{ab}$ in eq.(\ref{string}) and makes the
derivative dependence to $(d {\bar F})$, which vanishes.
Thus to one loop correction, the exponent in eq.(\ref{bulk}) 
containing $I_3^b\ [\ G(\t,\t)\ ]$ does
not essentially contribute to the result.{\footnote{Without an off-shell
extension of the amplitude, the integral $I_3^b\left [\ G(\t,\t )\ \right ]$
over the non-constant modes can also
be shown to vanish for a particular choice of
boundary condition.}}
On the other hand, the divergence in the Neumann propagator
at the boundary $G(\t,\t')$ can be seen to be 
absorbed by the re-normalization of the D-string coordinates which we would
discuss in the next section 3.4. 

\sp
\hs
Now from the expression (\ref{bulk}), one can re-write 
the disk amplitude (\ref{path})
as a path-integral over the boundary fields (say $\cz_{\pr\S}$)
consisting of auxiliary fields
$\nu_a(\t )$ as well as the normal coordinates $\z_a(\t )$. It reduces to

\be
\cs_{\rm eff}(f,A,B) \ \equiv \ {1\over{g_s}}\ \int \ dt \ d\s 
\ {\sqrt{{h \ + \ B \ + \ {\bar F}}\over{h}}} \ \ \cdot \cz_{\pr\S}
\ , \label{result}
\ee
where $\cz_{\pr\S}$ contains the boundary path-integrals involving the
auxillary fields $\nu_a(\t )$ and the normal coordinates $\z^a(\t )$.

\subsection{Evaluation of the boundary integrals}

As a second step, we perform the path-integral over the boundary
fields: Lagrange multiplier fields $\nu_a({\t})$ 
and the normal coordinates $\z_a(\t )$. Since the path-integral over
the Lagrange multiplier fields is a Gaussian, it is straightforward to
perform the integration.
However, the integral over the normal coordinates contains a quartic
interaction at the boundary.{\footnote{The details of the computations 
can be obtained from ref.\cite{karkazama} by generalizing the ${\bar F}_{ab}$
there to the invariant combination ${(B+\bar F)}_{ab}$. 
In addition, one needs to
switch off the tachyon vertex there.}} 
By performing the path-integration, the total contribution from the
boundary becomes

\be
\cz_{\pr\S} \ \equiv \ {1\over{\a'}} {\sqrt{-h}} \
\left ( \ 1 \ - \
\a' \ h \left [ h- \left ( B + \bar F \right ) \right ]^{-1} 
\cn_{(a)}^2\ \n_{aa} \
K^{\l}_{aa} \ K_{\l ab} \ h^{ab} \ \ln \eps \ \right )\ .\label{bound}
\ee
Now, the disk amplitude (\ref{result}) can be obtained 
by taking into account the bulk (Jacobian)
along with the boundary (\ref{bound}) contributions. To $\co(\a')$, we find

\bea
\cs_{\rm eff}(f,A,B) \ = \ 
T_D\ \int && dt\ d\s \ {\sqrt{- \Big ( \ h + B + {\bar F} \ \Big )}}
\nonumber \\
&& \cdot \left ( \ 1 \ - \ \a' 
\ h \left [ h - \left ( B + \bar F \right )\right ]^{-1}
\cn_{(a)}^2
\ \n_{aa} \ K^{\l}_{aa} \ K_{\l ab}
h^{ab} \ \ln \eps \ \right ) \ , \label{evaluation}
\eea
where $T_D = 1/(g_s \a' )$ denotes the D-string tenson. The leading term
(\ref{evaluation}) is precisely the generalized
Dirac-Born-Infeld action \cite{leigh} for a D-string in curved backgrounds.
The sub-leading term is the correction only due to the boundary interactions
and contains a divergence ($\ln \eps$). This indicates that the extrinsic
curvature on the D-string manifold is associated with the divergence.
However, the divergent correction can be
completely absorbed by a re-normalization of the world-sheet
fields $i.e. \ h_{ab} \ra h^R_{ab}$ and $ \ B_{ab} \ra B^R_{ab}$.
This in turn can be seen to be the shift $\d_af^{\mu}_R$ for a transformation
$\ f^{\mu} \ra f^{\mu}_R$ and becomes

\be
\d_af_R^{\mu} \ = \ - \ \a'\ \left ( {{h_R + B_R + \bar F}\over{h_R - \left (
B_R + \bar F \right )}}\right ) \ \n_{aa} \ \cn_{(a)}^2 
\ K^{\mu}_{aa} \ \ln \eps 
\ . \label{ren}
\ee
Then the renormalized form of the amplitude can be expressed 
exactly by the generalized Dirac-Born-Infeld action. To $\co(\a' )${\footnote
{Henceforth, we omit the subscript or superscript `R' denoting
the renormalized D-string.}:

\be
\cs_{\rm eff}(f,A,B) \ = \ 
T_{D} \ \int \ dt \ d\s \ 
{\sqrt{- \Big ( \ h \ + \ B \ + \ {\bar F} \ \Big )}} \ . \label{dbi}
\ee

\sp
\hs
The renormalized DBI action obtained (\ref{dbi}) would receive corrections
(in $\a'$) from the higher orders 
in the derivative expansion of the D-string coordinates $f^{\mu}(t,\s )$.
On the other hand, the WZ action remains invariant under
the transformation (\ref{ren}), which signifies its exact nature.
The complete dynamics of D-string can
be described by the combination of renormalized DBI and WZ actions.

\section{Aspects of D-string in Quantum Dynamics}

\subsection{Quantization of collective coordinates}
 
\hs
In this section, we discuss the covariant quantization in the $U(1)$
gauge sector of D-string collective coordinates $A_a$.
We begin with the D-string in type IIB string backgrounds
described by the renormalized DBI and WZ actions. The DBI action obtained
(\ref{dbi}) describes the dynamics of D-string and the WZ action (\ref{wz})
gives rise to the RR charges to the D-string.
As discussed before, in presence of the background fields, 
the lower form in the RR sector couples to the $U(1)$ gauge invariant
field strength. Thus one needs to take into account the WZ term while
deriving the quantum dynamics of D-string.

\sp
\hs
With the set up, we begin by writing the quantum amplitude for the
collective coordinates of D-string ($\ f^{\mu} , A_a \ $ )

\be
\cz \ = \ \int \ {\cd }f^{\mu} \ {\cd }A_a \ \cdot \exp \left ( 
- \ \int \ dt \ d\s \ {\cal L} \ \right ) \ ,
\label{dbiwz}
\ee
where ${\cal L}$ is the effective Lagrangian describing a curved
D-string in the arbitrary backgrounds
 
\be
{\cal L} \ =  \ T_D  \
{\sqrt{- \Big ( \ h \ + \ B \ + \ {\bar F} \ \Big )}} \ + \ {Q\over2} \
{\ce }^{ab} \ \Big (\ C_{ab} + (B_{ab} + {\bar F}_{ab}) \ \chi \ \Big )\ .
\ee
Now we would like to present
the canonical anlysis for an arbitrary D-string with its collective coordinates
$(f^{\mu},A_a)$. The canonical conjugate momenta are $p_{\mu}$ and 
$E^a$ (conjugate electric field) respectively. They can be expressed as

\be
p_{\mu} \ = \ {{\pr {\cal L}}\over{\pr {\dot f}^{\mu}}} 
\ \ \ {\rm and}\ \ \ \ \ E^a \ = \ {{\pr {\cal L}}\over{\pr {\dot A}_a}} \ .
\label{momentum}
\ee
Explicitly 
\bea
p_{\mu} &=& {{-\ T_D}\over{\sqrt {-(h+ B +\bar F)}}} \left (
G_{\mu\nu}\dot f^{\nu}
h_{11} - G_{\mu\nu} f'^{\nu}h_{01} + (B+\bar F)^{1\over2}
 \ B_{\mu\nu}{f'}^{\nu}\right )
+ \ Q\ C_{\mu\nu}{f'}^{\nu}\nonumber \\ 
{\rm and} \ \ \
E^1 &=& -\ T_D \ \left ( - \ {{B + \bar F}\over{h\ +\ B \ +\ {\bar F}}} 
\right )^{1\over2} \ + \ Q \ \chi \ , \nonumber \\
&\equiv & {\tilde E} \ + \ Q \ \chi \ ,
\label{efield}
\eea
where $\tilde E$ is the usual electric field due to the KR two-form.
It is interesting to note that the conjugate electric field $E=(E^0,E^1)$
receives correction from the RR lower form: $\chi$. However for a vanishing
KR background or the $U(1)$ gauge field, the electric field $E$ vanishes.
The consistency condition $E^0 \ = \ 0$ 
obtained from eq.(\ref{momentum}) 
is a primary constraint and then the Gauss law constraint becomes
 $\pr_1E^1 \ = \ 0 $ and appears as a secondary constraint in the theory. This
implies the static condition for the electric field $\tilde E(t)$ and the 
RR zero-form $\chi (t)$. Furthermore, the canonical momenta
($p_{\mu},E^a$) in the theory give rise to another set of primary constraints

\be
L_{\pm} \ = \ {1\over2} \ \left (\ p \ \pm \ T_{\tilde E}\ f' \ \right )^2 \ 
= \ 0 \ ,
\label{virasoro}
\ee
where $T_{\tilde E}^2 = {T_D}^2 + {\tilde E}^2$ and shall be identified with 
the modified
D-string tension in presence of ( background fields ) electric field 
$\tilde E$. Furthermore, the constraints $L_{\pm}$ can be shown to 
be closed under the left and
right Virasoro algebras. One can then construct the BRST charge (nilpotent)
of the D-string in presence of background fields from the covariant set of
constraints ($E^0,L_{+},L_{-}$) and obtains the covariant gauge fixed action.
The details of the construction can be
obtained following ref.\cite{hatsuda}. 
Finally, one finds a path-integral over all the phase space variables 
$(\ f^{\mu},A_a,p_{\mu},E^a \ )$. Performing the path-integration over the
conjugate momentum $p_{\mu}$, we obtain

\bea
\cz \ = \ \int \ {\cd }f^{\mu} \ {\cd }A_a \ {\cd }{E}^1 
\ \cdot \exp \Big ( \ \int \ dt \ d\s && 
\Big [ - \ T_{\tilde E}
\ {\sqrt{- h}}  \
- \ Q \ C_{01} \ -\ Q\ E^1 \ B_{01} \ \chi \nonumber \\
&&{}\ \ \  + \ E^1\ \pr_0 A_1 \ + \ A_0\ \pr_1 E^1 
\ \Big ] \ \Big )\ . \label{phase}
\eea
By performing the path-integration over the electric field $E^1 \equiv E$,
one goes back to the amplitude (\ref{dbiwz}) describing the quantum
dynamics of D-string in presence of background fields. 

\sp
\hs
On the other hand, the path-integration over the gauge field $A_a$ 
in the action (\ref{phase}) can be performed and the integral becomes

\bea
\cz \ = \ \int \ \cd f^{\mu} \ \ \cd E^1 \ \d (\pr_0E^1 ) \ \d (\pr_1E^1) \ 
&& \cdot \exp \Big ( \ \int \  dt \ d\s \Big [-\ T_{\tilde E} \ {\sqrt{-h}}
\nonumber \\
&& {}\ \ \ \  \ \  - \ Q \ C_{01} \ - \ 
Q \ E^1\ B_{01}\ \chi \ \Big ] \ \Big ) \ .
\eea
In addition, the electric field $E\ ( =\ m)$ is quantized (m denotes integers), 
since its conjugate
field $A_a$ is defined on a Wilson loop. Then the path-integral over the
electric field $E^1$ reduces to that over its zero mode and the amplitude
takes the form:

\be
\cz \ = \ {\sum }_{m}
\ \int \ \cd f^{\mu} \ \cdot\exp \left (\ \int \ dt \ d\s \
\left [ \ - \ T_{\tilde E} \ {\sqrt{- h}} \ - \ Q \ C_{01} 
\ - \ m\ Q\ B_{01} \ \chi \ \right ] \ \right ) \ , \label{eldstring}
\ee
Now the D-string quantum dynamics can be read off from the above path-integral.
The first term in the action there in eq.(\ref{eldstring}) 
is precisely the Nambu-Goto action 
describing the dynamics of a D-string with a modified
string tension $T_{\tilde E}$ and is identical to that of the 
Green-Schwarz action.
The second and third terms correspond to the WZ action, reflecting
the RR charges (as expected) for a D-string in general backgrounds. 
Following the standard method for a F-string, the Polyakov form of the action
for a D-string can be obtained as

\be
S_{\rm D-string} \ = \
{{T_{\tilde E}}\over2} \ \int \ dt \ d\s \ {\sqrt{-\ h_{({\rm int})}}} 
\ h_{int}^{ab} \ \pr_a 
f^{\mu} \pr_b f^{\nu} \ G_{\mu\nu} \ 
+ \ {Q\over2}\ \int \ dt \ d\s \ C \ \cdot\exp{\left (\ m\ B \ \right ) }\ ,
\label{poly}
\ee
where $h_{(\rm int)}$ is the intrinsic metric on the D-string world-sheet
and the WZ action is expressed schematically with respect to the induced
forms on the D-string. 
Here, one can note that the coefficients of the first and second terms are 
different for a D-string
unlike that for a Green-Schwarz string.  In a conformal gauge 
$h_{({\rm int})ab}=
e^{\tilde\b } \ \n_{ab}$ ($\tilde\b $ : D-string Livouille field),
the D-string action (\ref{poly}) reduces to

\be
S_{\rm D-string} \ = \ {{T_{\tilde E}}\over2} \ \int \ dt \ d\s \
G_{\mu\nu}\  \pr_a f^{\mu} \pr_a f^{\nu} \ + \ 
{Q\over2}\ \int \ dt \ d\s \  {\ce }^{ab}\ \ C_{ab} \ \cdot\exp{\left ( \ m\ B
\right )} \ .
\label{sigma}
\ee
This represents the Polyakov form of the action for a D-string in 
a general curved background fields.
 
\subsection{D-string bound states}

Now we briefly analyze the D-string tension $T_{\tilde E}$
obtained in presence of the quantized electric flux `$m$' due to the
fundamental strings. From eq.(\ref{efield}), the
electric field can be seen to receive a shift from the RR lower
form $\chi (t)$ present on the world-sheet of D-string. This implies that the
D-string tension in presence of electric field couples to a RR zero-form.
The modified D-string tension in the fundamental string units ($2\pi\a' =1$)
can be given
\be
T_{\tilde E} \ = \ {\sqrt{\ \left (\ m\ - Q\ \chi \ \right )^2 \ 
+ \ {1\over{{g_s}^2}} }}\ .
\label{tension}
\ee
Now the D-string tension $T_{\tilde E}$ for a stationary axion $\chi $,
can be identified with the tension
$T_{(\ m,1 \ )}$ in string units of an
(m,1) string. Thus it is a truly non-perturbative bound state 
\cite{witbound} of a D-string and the electric flux (`$m$' units) of F-string.
Since the origin of electric flux  is in 
the $U(1)$ gauge invariant
combination of $(B+\bar F)_{ab}$ in the NS-NS sector, the above bound state
is natural and inherent to a D-string in type IIB theory. However in the
week coupling limit, the D-string becomes infinitely heavy
and the F-strings almost lose their energy in the process of binding to a
D-string. On the other hand, in the strong coupling limit the D-string
becomes light and can be described in terms of Green-Schwarz string. 
Furthermore, in absence of electric field $E$, the D-string
is precisely described by the Nambu-Goto action. This is the case for a
single D-string in type I theory, since under the world-sheet parity invariance
the continuous $U(1)$ gauge group is projected out leaving its discrete 
subgroup $Z_2$. In such a case, the D-string states may be identified with
the states of Green-Schwarz string.

\subsection{Boosted D-string}

In this section, we would like to analyze the boundary conditions 
obtained in eq.(\ref{bcond}) for the D-string quantum dynamics.
We re-write the modified Neumann boundary conditions
in terms of components

\be
\pmatrix { {\pr_n} & {i\ {m\over{T_{\tilde E}}}\ \prt } 
\cr {-i \ {{m}\over{T_{\tilde E}}} \ \prt }
& {\pr_n} \cr } \ \pmatrix { {\r_0} \cr {\r_1} \cr } \ = \ 0 \ .
\ee
Now considering a boost ($R$-matrix) on the F-string coordinates 
${\tilde\r }_A \ = \ R_A^B \ \r_B$ for ($R_{\a}^{\b }= 1,1,\dots , 1 $),
it is straightforward to see that the modified Neumann boundary conditions 
become 
\be
\pr_{m_{\pm}} \tilde\r_a \ = 0 \ ,
\label{mixed}\ee
where $\pr_{m_{\pm}} = \pr_n \pm i {{m}\over{T_{\tilde E}}} \prt $. 
The boundary
conditions (\ref{mixed}) are the usual Neumann boundary conditions in the
(transformed)
tangential directions at the cost of a boost which is essentially due to 
the flux of F-string giving rise to the electric field $\tilde E$. However, the 
boost is quantized while describing the quantum dynamics 
of D-string. 

\sp
\hs
On the other hand, in the semi-classical picture of D-string
($i.e.$ before quantization of gauge field) the boost can be seen as 
the natural one. There in the the boosted frame, the Neumann propagator
matrix $G_{ab}(z,z')$ can be seen to absorb all the background fields.
Following similar analysis as shown in this work, one can  
obtain Nambu-Goto 
type action along with WZ action 
for the boosted D-string on the non-zero mode sector. However the
zero-mode in the theory would contribute towards the non-commutative
nature. This implies
that the D-string in presence of arbitrary background 
fields can be viewed as a boosted D-string 
which is essentially due to the F-string flux.

\subsection{Fluctuations in presence of backgrounds}

\hs
In this section, we shall analyze the fluctuations on the D-string associated
with the ends of open string lying on its world-sheet. The ends of the open
string carries $U(1)$ gauge field, which has been seen as a consequence
of the KR two-form potential. In fact, these background fields couple to
the appropriate RR forms and carry RR charge due to the lower form.
Let us begin with the WZ action obtained after re-normalization of D-string
coordinates $f^{\mu}(t,\s )$. One can rewrite the WZ action schematically from 
eq.(\ref{sigma}) for an Euclidean curved D-string :

\be
S_{WZ}(f,C) \ = \ i \ Q \ \int_{S_2} \ 
d^2\s \ \Big ( \ C_2 \ + \ m\  B_2 \wedge \chi \ \Big )
\ . \label{explicit}
\ee
In this case `$m$' denotes the quantized magnetic flux corresponding to the
fundamental string. Now let us go back to the WZ action in a three dimensional
compact manifold $S_3$ with a boundary $\pr S_3=S_2$ describing the
world-sheet of D-string. Since the induced KR two-form 
($B_{ab}= {\sqrt{- h}}\ \ce_{ab}$ )
is covariantly constant for a D-string, the WZ action becomes

\be
S_{WZ}(f,C) \ = \ i \ Q \ \int_{S_3} \
d^3\s \ \Big ( \ H_3^{RR} \ + \ m\ B_2  
\wedge H_1^{RR}
\Big ) \ , \label{RRstrength}
\ee
where $H_1^{RR}$ denotes the RR one-form 
field strength induced on the compact manifold $S_3$.
The expression (\ref{explicit}) for the WZ action 
can be better understood by allowing the
F-string to wrap around a compact two-cycle $S_2$. Then the winding number
becomes

\be
w\ = \ {1\over{2\pi}} \ \int_{S_2} \ d^2\s \ B_2
\ .
\ee
Finally, the WZ action (\ref{RRstrength}) for an Euclidean D-string in 
presence of non-trivial background fields can be given

\be
S_{WZ}(f,C) \ \equiv \ iQ \ \int_{S_2} \ 
d^2\s \ C_{2} \ + \ i \ m' \ {Q_i} \ \left [ \
\chi\ \right ]_{\rm fixed} \ , \label{instanton}
\ee
where $Q_i$ denotes the chrage density of D-instanton \cite{green,gibbons} 
and is associated with the pseudo-scalar axion $\chi $ induced on the D-string.
Here $m'= (m w)$ denotes the discrete magnetic flux of fundamental
strings and $\left [\chi \right ]_{\rm fixed}$ corresponds to the
position of a D-instanton. This implies that,
in presence of an arbitrary KR two-form, the D-string can 
also be seen to
possess D-instanton charge. There are
`$m'$' units of D-instantons 
acting as a source for the axion
whose positions are fixed in the Euclidean space. 
It is clear that, the additional non-perturbative contributions are 
essentially due to the KR two-form or the $U(1)$ gauge field.
If D-instanton charge is consistent on a D-string,
then their configurations must also be supported by 
the appropriate boundary conditions.
We shall see that, it is indeed the case. The boundary conditions for a
D-instanton are defined by the Dirichlet conditions $\r_{A'}(\t)=0$ for
$(A'\equiv (a',\a)= 26,1,2,\dots , 25)$ representing its position.
On the other hand, for an Euclidean D-string the boundary conditions are
described by the Neumann $\pr_n \r_{a'}= 0$ for $(a'=26,1)$
and the rest by Dirichlet conditions $\r_{\a}(\t ) =0$ in the transverse 
directions. Since the zero-modes are separated out from the string coordinates,
one can re-write the above boundary conditions for the D-instanton
\be
\prt \r_{a'} (\t)\ = \ 0 \ \ \ \ {\rm and }\ \ \ \ \ \ \r_{\a}(\t) \ = \ 0 \ .
\ee
Now let us consider `$n$' units of D-instantons superpose on
the D-string configuration.
By recalling the boundary conditions obtained in eq.(\ref{bcond}),
it can be seen to be compatible with the above
bound state construction for non-vanishing background fields 
$\Big ( B_{ab}+{\bar F}_{ab}\ \neq 0 \Big )$ and $n=m'$. 
However in the Lorentzian space 
such a bound state of D-string and D-instanton can not be addressed. 
Nevertheless, such point configurations (D-instanton like) are non-localized 
due to the zero modes on the D-string in presence of NS-NS background fields. 
Since the background fields effectively interact at the boundary of the 
D-string, one can find that the fluctuations on the D-string are associated
with such D-instanton like point configurations.

\section{Discussion}

\hs
To summarize, the non linear dynamics of an arbitrary D-string in
the bosonic sector of type IIB superstring theory have been reviewed 
in the present work using a path-integral formalism which is a
natural approach. In the process,
we have generalized our earlier formulation for a D-string with type I 
background field ( $U(1)$ gauge field ) by incorporating the general
curved background fields from the NS-NS and RR sectors. Since D-string
is intrinsic to type IIB theory (for reasons discussed in the text),
the path-integral formulation for D-string in type IIB backgrounds may
be considered as an important step towards our understanding of the 
D-string quantum dynamics. It has been shown that the effective dynamics
of D-string is best approximated by the generalized Dirac-Born-Infeld action
in the bosonic open string theory where as the RR charges are described
exactly by the Wess-Zumino action. In presence of the NS-NS two-form
field, the D-string receives an additional RR charge and we have performed 
covariant quantization by taking into account the above coupling. We have
out-lined some of the noble aspects of D-string quantum dynamics in the
path-integral formulation.

\sp
\hs
As a step ahead, an important issue is the supersymmetry in the
path-integral formalism for an arbitrary D-string $f^{\mu}(t,\s )$. To 
our understanding, this problem is difficult due to non-linear
dynamics associated with the fermionic
partners of $f^{\mu}(t,\s )$ though the F-string fermions may be handled.
It is also interesting to formulate
the path-integral to describe the non-linear dynamics of D-brane in the open 
string language.
This is a non-trivial generalization from D-string to arbitrary D-brane 
mainly due to the higher dimensional D-brane 
manifold and higher forms in RR sector. Presently, it is under study.

\sp
\hs In comparison to the Green-Schwarz
fundamental string, the D-string can be seen to contain extra degrees of 
freedom. These are essentially due to the $U(1)$ gauge field associated with
the ends of the open string lying on the world-sheet $f^{\mu}(t,\s )$. The
extra degrees are responsible for a highly non-linear dynamics of D-string
and lead to the effective description of D-string in contrast to the 
exact dynamics as known
for a Green-Schwarz string. Integration of these extra degrees, gives rise
to a discrete electric field and may require a deeper understanding.

\sp
\sp
\sp

{\large \bf Acknowledgments:} 
\sp

\hs I gratefully acknowledge Y. Kazama for 
collaboration in ref.\cite{karkazama} and various useful discussions then.
I would like to thank the members of high energy physics group at Institute
of Physics, Bhubaneswar for many illuminating discussions, 
on a part of the present
work, during a short visit. Also, I am grateful to the members of string theory
group in the Institute here for discussions during an informal presentation of
this work. Especially, I would like to thank M. Cederwall and M. Henningson 
for their comments and suggestions. This work is supported by Swedish Natural 
Science Research Council.

\def\anp{Ann. of Phys.}
\def\prl{Phys. Rev. Lett.}
\def\prd#1{{Phys. Rev.} {\bf D#1}}
\def\plb#1{{Phys. Lett.} {\bf B#1}}
\def\npb#1{{Nucl. Phys.} {\bf B#1}}
\def\mpl#1{{Mod. Phys. Lett} {\bf A#1}}
\def\ijmpa#1{{Int. J. Mod. Phys.} {\bf A#1}}
\def\rmp#1{{Rev. Mod. Phys.} {\bf 68#1}}

\vfil\eject

\end{document}